\documentstyle[prl,aps,psfig]{revtex}
\input epsf
\begin{document}

\draft

\preprint{\vbox{
\hbox{CWRU-P13-1998}
}}

\title{On Random Bubble Lattices}

\author
{
Andrew A. de Laix$^{{\dag}{\ddag}}$ and Tanmay Vachaspati$^{\ddag}$
}
\address{
$^{\dag}$Wolfram Research, Inc., 100 Trade Center Drive, 
Champaign IL, 61820. \\ 
$^{\ddag}$Physics Department, Case Western Reserve University,
Cleveland OH 44106-7079.
}

\date{\today}

\twocolumn[
\maketitle

\begin{abstract}
\widetext

We study random bubble lattices which can be produced by processes
such as first order phase transitions, and derive characteristics that 
are important for understanding the percolation of distinct varieties
of bubbles. The results are relevant to the formation of topological
defects as they show that infinite domain walls and strings will be 
produced during appropriate first order transitions, and that the most 
suitable regular lattice to study defect formation in three dimensions 
is a face centered cubic lattice. Another application of our work is to 
the distribution of voids in the large-scale structure of the universe.
We argue that the present universe is more akin to a system undergoing 
a first-order phase transition than to one that is crystallizing, as
is implicit in the Voronoi foam description. Based on the picture of a 
bubbly universe, we predict a mean coordination number for the voids of 
13.4. The mean coordination number may also be used as a tool to 
distinguish between different scenarios for structure formation.

\end{abstract}

\pacs{}

]

\narrowtext

The formation of topological defects has mostly been studied by
numerical techniques on a regular lattice
\cite{tvav,robinsonyates,rsav}; yet it is known that the
numerical results for the statistical properties of topological
defects depend on the lattice that is chosen for performing
the simulations \cite{rsjf}. 
Our primary objective here is to
consider defect formation during first order phase transitions in the
continuum (also see \cite{jbtktvav,jb}) and to determine if infinite 
domain walls and strings will be produced.
The results we obtain,
however, are of wider applicability, since they apply to any process
which leads to a random lattice. An example is the growth of universal
large-scale structure, which leads to an observed distribution of voids
in galaxy surveys.

First-order phase transitions proceed by the nucleation of bubbles of
the low temperature phase in a background of the high temperature
phase.  The bubbles then grow, collide, and coalesce, eventually
filling space with the low temperature phase. In a variety of
circumstances, the low temperature phase is not unique. Here we
will first consider the case where there are two low temperature
phases, which we call plus ($+$) and minus ($-$). The interface
between two spatial regions containing different phases is called a ``domain
wall''. We will be interested in determining if the $+$ and $-$ phases
percolate ({\it i.e.} form infinite clusters) after the phase transition. 
Given that the probability of a bubble being in the plus phase is $p$, there
exists a critical probability $p_c$ such that the + phase percolates
only if $p > p_c$. Our goal is to determine $p_c$.  If $p_c$ is found
to be less than 0.5, then a range of $p$ exists for which both the $+$
and $-$ phases will percolate and, in this case, infinite domain walls
will be formed \cite{tvtrieste}. We will also consider the formation
of strings on the bubble lattice following the algorithm described in
Ref. \cite{tvav}. Here we will find that strings percolate on the bubble
lattice and that there is an infinite string component to the 
network that is somewhat larger than that found in Ref. \cite{tvav}.

Let us begin by studying the structure of the random bubble lattice
that is produced during a first order phase transition and later
discussing percolation on this lattice. We write the bubble nucleation rate 
per unit volume as $\Gamma$, and we assume that the bubble
walls expand at constant speed $v$. From these quantities we can define 
a length scale $\xi$ and a time scale $\tau$ by:
\begin{equation}
\xi = \biggr ( {v \over \Gamma} \biggr )^{1/4} ~ , \ \ 
\tau = {1 \over {(v^3 \Gamma )^{1/4}} } ~ ,
\end{equation}
where the exponents have been shown for bubbles in three dimensions. 
By rescaling all lengths (such as bubble radii) and all times by $\xi$
and $\tau$ respectively, the dependence of the problem on $\Gamma$ and
$v$ is eliminated. Therefore dimensionful quantities such as the number
density of bubbles of a given size can be rescaled to a universal distribution, 
and dimensionless quantities, such as the the critical percolation probability, 
will be independent of $\Gamma$ and $v$. 

The scaling argument given above relies on the absence of any other
length or time scales in the problem. Potentially such a scale is provided by
$R_0$, the size of bubbles at nucleation, and our assumption is that
$R_0 << \xi$. Also, note that we have taken all bubbles to expand at
the same velocity $v$. This is justified if the low temperature phases
within the bubbles are degenerate. If this degeneracy is lifted, different
bubbles can expand at different velocities and this may result in lattices
with varying properties. We are primarily interested in the exactly
degenerate case which is relevant to the formation of topological defects.

In a computer simulation,
there are two other scales that enter. These are the size of the 
simulation and the time over which the process is studied. If the
size of the simulation is very large compared to $\xi$, the length scale is
effectively infinite and does not play a role in the properties of the
bubble lattice. Also, we are only interested in the bubble statistics once 
the bubbles fill the simulation volume and the phase transition is 
complete. At this stage, the properties of the bubble lattice are
fixed and further evolution does not play a role. Hence, if we observe
the lattice at any time greater than the phase transition completion
time, the observation time will not enter the properties of the bubble
lattice. We have verified numerically that quantities such as the size
distribution of bubbles are universal. Hence we can simulate 
the random bubble lattice for any convenient choice of parameters.

We have simulated the nucleation and growth of bubbles leading to the 
completion of the phase transition following the scheme described in 
Ref. \cite{jbtktvav}. There are two ways to view this scheme. The first is
a dynamic view where, as time proceeds, the number of nucleation sites 
are chosen from a Poisson distribution, bubbles keep growing and 
colliding until they fill space. The second equivalent viewpoint is 
static and more convenient for simulations. A certain number of
spheres whose centers and radii are drawn from uniform distributions
are placed in the simulation box. This corresponds to a snapshot of the
bubble distribution. If the number of spheres that are laid down is
large, they will fill space and the snapshot would be at a time after
the phase transition has completed.

It is worth comparing the present model with currently existing models 
of froth.  The main distinguishing feature is that the bubbles continue to 
grow even after they collide. This is in sharp distinction with the models 
used in crystal growth such as the Voronoi and the Johnson-Mehl models. 
In these models, crystals nucleate randomly inside a volume, 
grow and then, once they
meet a neighboring crystal, stop growing in the direction of that
neighbor. (In the Voronoi model, all crystals are nucleated at one
instant while in the Johnson-Mehl model, they can nucleate at different
times.) This difference between the phase transition model and the
Voronoi type models is significant and the resulting lattices have
different properties.
Another model considered in the literature is called a
``Laguerre froth''. Here the snapshot of the domains corresponds to 
a horizontal slice of a mountain range in which each mountain 
is a paraboloid. The circles of intersection of the plane and the
paraboloids define the Laguerre froth \cite{rivier}. In terms of bubbles, 
this means that the bubble walls move with a velocity that is proportional 
to $\sqrt{t-t_0}$ where $t-t_0$ is the time elapsed since nucleation. 
Such a model in two dimensions was studied by numerical
methods in Ref. \cite{telley}. If the paraboloids are replaced by cones, 
the model comes closer to the present one. Such a model has not been 
analyzed previously in any number of dimensions.

A feature of our model of the first order phase transition is that
bubbles cannot nucleate within already existing bubbles. This is
appropriate to the case where the phases existing within bubbles are
degenerate or nearly degenerate. However, in cases where a variety of 
non-degenerate bubbles can exist (for example, if the system has metastable
vacuua), this assumption may have to be relaxed \cite{gleiseretal}. 

\begin{figure}[tbp]
\caption{\label{lattice}
A portion of the three dimensional dual bubble lattice.
}
\epsfxsize = \hsize \epsfbox{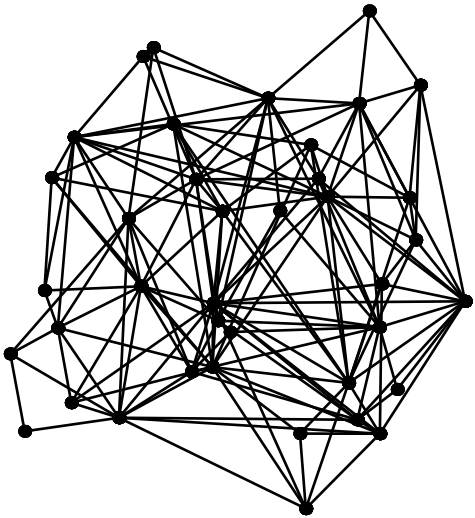}
\end{figure}

We construct the three dimensional (dual) bubble lattice by 
connecting the centers of bubbles that have collided (Fig. \ref{lattice}). 
The bubble lattice is almost fully triangulated though some violations of 
triangulation can occur. For example, if a tiny bubble gets surrounded by 
two large bubbles, the center of the tiny bubble will only be
connected to the centers of the two surrounding bubbles and this
can lead to plaquettes on the lattice that are not triangular.
The characteristics of this bubble
lattice hold the key to the percolation of phases and the formation of
topological defects. In particular, the average number of
vertices to which any vertex is connected is expected to play a crucial
role. This number is called the ``mean coordination number'' of the lattice,
and we now determine this quantity analytically.

First we consider the two dimensional case. We denote the number of points
in the lattice by $P$, the number of edges by $E$ and the number of
faces by $F$. Then the Euler-Poincar\'e formula \cite{nashsen} tells us
\begin{equation}
\chi = P - E + F
\label{eulerpoincare2d}
\end{equation}
where, $\chi$ is the Euler character of the lattice and is related to the
number of holes in the lattice (genus).
In our case, the lattice covers a plane which we can compactify in some
way, say by imposing periodic boundary conditions. Then $\chi$ is the
genus of the compact two dimensional surface. For us it will only be
important that $\chi = O(1)$. Next, if ${\bar z}$ is the (average) coordination 
number, we can see that 
$$
E = {{\bar z} \over 2} P ~ ,
$$
since, a given point is connected to ${\bar z}$ other points but each edge is
bounded by two points. Also,
$$
F = {2 \over 3} E ~ ,
$$
since each line separates 2 faces but then each face is bounded by 3 lines.
Now, using (\ref{eulerpoincare2d}) gives
$$
1 - {{\bar z} \over 2} + {{\bar z} \over 3} = {\chi \over P} \simeq 0 ~ ,
$$
since $P$ is assumed to be very large. Therefore, in two dimensions,
${\bar z}=6$, a result that first appeared in the botanical 
literature \cite{botany,rivier}. 

In three dimensions the analysis to evaluate ${\bar z}$ is somewhat more
complicated. The Euler-Poincar\'e formula now says
\begin{equation}
\chi = P - E + F -V ~ ,
\label{eulerpoincare3d}
\end{equation}
where $V$ is the number of volumes in the 
lattice. Now, in addition to the usual coordination number ${\bar z}$, we 
also need to define a ``mean face coordination number'' ${\bar y}$ which 
counts the average number of faces sharing a common edge. In terms of 
${\bar y}$ and ${\bar z}$, the relations between the 
various quantities for a triangulated three dimensional lattice are:
\begin{equation}
E = {{\bar z} \over 2} P ~ , \ \ 
F = {{\bar y} \over 3} E ~ , \ \ 
V = {2 \over 4} F  ~ ,
\label{reltaions}
\end{equation}
where the first equation is as in two dimensions, the second equation
follows from the definition of ${\bar y}$ and the fact that the lattice is
triangulated, and the last relation follows because a face separates
two volumes and a volume is bounded by four faces that form a tetrahedron.
Inserting these relations in (\ref{eulerpoincare3d}) leads to:
\begin{equation}
{\bar z} = {{12} \over {6-{\bar y}}}  ~ ,
\label{yzrelation}
\end{equation}
where, as before, we assume that $P$ is very large and ignore the
$\chi /P$ term.
Note that the relation between ${\bar y}$ and ${\bar z}$ is purely topological and
will hold for any triangulated lattice. 

We now want to estimate ${\bar y}$. For this we work in a ``mean
field'' approximation where we assume that the edge lengths 
are fixed. We consider two vertices $A$ and $B$ separated
by a unit distance. Now we wish to find the number of points that can
be connected to both $A$ and $B$, subject to the constraint that
connected points are at unit distance from each other. This will give
the (average) number of faces that share the edge from $A$ to $B$ and
hence will be the face coordination number ${\bar y}$.  Let us choose
$A$ to be at the center of a sphere of unit radius and $B$ to be at
the North pole. Then the additional points $P_1$,...,$P_y$, have to
lie on the circle at latitude 60 degrees to satisfy the distance
constraint. Then one finds that the azimuthal angular separation of
two sequential points $P_i$ and $P_{i+1}$ is 70.5 degrees. Therefore
\begin{equation}
{\bar y} = {{360} \over {70.5}} = 5.1  ,
\label{yestimate}
\end{equation}
which then leads to \cite{coxeter}
\begin{equation}
{\bar z} = 13.4 ~ .
\label{zestimate}
\end{equation}

It is worth noting the ingredients that have entered into the analytic
estimate of ${\bar z}$. The relation (\ref{yzrelation}) is a
topological statement about the lattice, but the estimate for ${\bar
y}$ is geometric, depending on the assumption that the 
edges have fixed length. In principle, the edge lengths can 
fluctuate but our estimate for ${\bar y}$ will still 
be valid if the fluctuations average out. 

\begin{figure}[tbp]
\caption{\label{neighbor}
The coordination number relative frequency 
distribution for the three dimensional dual bubble lattice.
}
\epsfxsize = \hsize \epsfbox{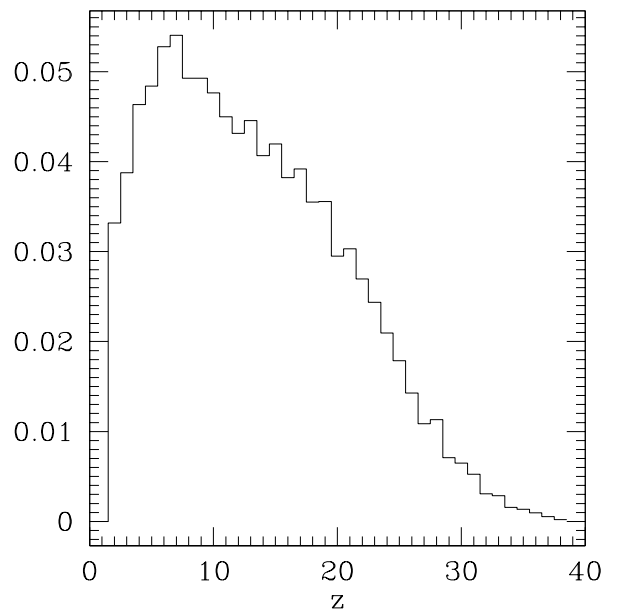}
\end{figure}

In Fig. \ref{neighbor} we show the distribution of coordination number
in our three dimensional simulations. The average coordination number
is found to be ${\bar z}=13.34 \pm 0.05$ and agrees quite closely with 
the mean field result. For comparison, Voronoi foam has ${\bar z} = 15.54$
and the Johnson-Mehl model has ${\bar z} > 13.28$ \cite{meijring}. 
The reason why ${\bar z}$ is larger in the Voronoi model is that, 
in this model, the cells stop growing on collision in the direction of
the collision, thus leading to anisotropic growth. It can be shown
that anisotropy of the cells leads to a higher value of ${\bar z}$ 
\cite{rivier}.

The mean value of $z$ is not a good characteristic of the distribution
of $z$ since the distribution is skewed (the modal value of $z$ is 7)
and it is of interest to characterize the entire distribution of $z$. 
In the literature on domain physics, 
attempts to derive the distribution of coordination number are often based 
on maximizing the ``entropy'' of the lattice subject to the constraints in 
the system. The expression used for the entropy is the one proposed by 
Shannon \cite{shannon,rivier2}. On employing this procedure, one finds 
an exponential fall-off of the distribution. The distribution shown in 
Fig. \ref{neighbor} also has an exponential fall-off:
\begin{equation}
f(z) \sim \exp [ -0.25 z ] \ , \ \ z  \gtrsim 20 \ .
\label{falloff}
\end{equation}

\begin{figure}[tbp]
\caption{\label{perc}
The zeroth, first and second moments of the cluster distribution function
versus the probability $p$ in three dimensions. 
}
\epsfxsize = \hsize \epsfbox{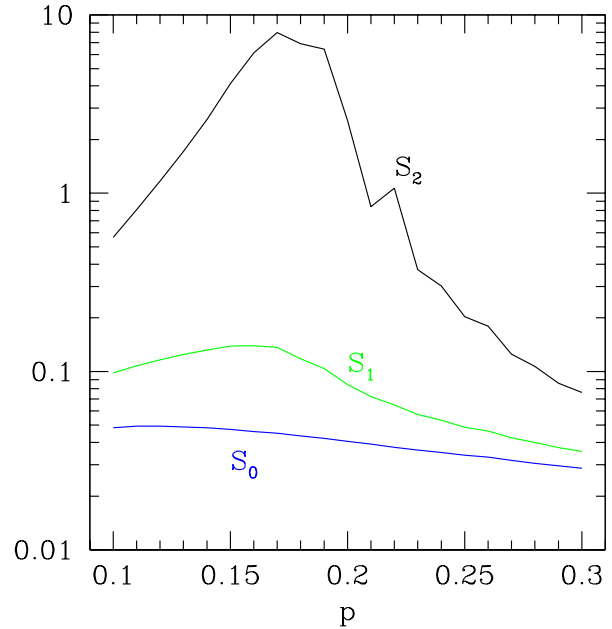}
\end{figure}

We now turn to the formation of defects on the bubble lattice. We 
put a + phase on a bubble with probability $p$ and a $-$ phase with 
probability $1-p$. We then find the size distribution of + clusters 
and calculate the moments of the cluster distribution function after 
removing the largest cluster from the distribution \cite{stauffer}. 
That is, we calculate:
\begin{equation}
S_l (p) = \sum_{s \ne s_{max}} s^l  n_s (p)
\label{momentsdefn}
\end{equation}
for $l=0,1,2,...$, where the sum is over cluster sizes ($s$) but does
not include the largest cluster size, and $n_s (p)$ is the number of
clusters of size $s$ divided by the total number of bubbles.  In
Fig. \ref{perc} we show the first three moments as a function of $p$,
where the turning point in $S_2$ marks the onset of percolation. To
understand this, first consider the behavior of the second moment for
small $p$. As we increase $p$, there are fewer + clusters (as seen
from the $S_0$ graph) probably due to mergers, but the merged cluster
sizes are bigger (as seen from the $S_1$ graph).  Since the second
moment places greater weight on the size of the cluster than on the
number density as compared to the lower moments, it grows for small $p$. 
For large $p$, however, as we
increase $p$ further, the additional + clusters join the largest
cluster of +'s and are not counted in the second moment. In fact, some
of the smaller clusters also merge with the largest cluster and get
removed from the sum in (\ref{momentsdefn}). This causes the second
moment to decrease at large $p$. Hence, the second moment has a
turning point and the location of this turning point at $p_c$ marks
the onset of percolation. In three dimensions we find $p_c =
0.17\pm 0.01$ (from Fig. \ref{perc}), which is well under 0.5, 
while in two dimensions we find $p_c =0.50 \pm 0.01$ which is
consistent with 0.50. (The two dimensional version of Fig. \ref{perc}
may be found in \cite{tvtrieste}.)

It is interesting to compare the critical probabilities we have found
with lattice based results for site percolation where the regular lattice 
has a coordination number close to that of the random bubble lattice. In
two dimensions a triangular lattice has ${\bar z}=6$ and $p_c
=0.5$. In three dimensions, a face centered cubic lattice has ${\bar
z}=12$ and $p_c = 0.198$ \cite{stauffer}.  These values of the
critical probabilities are fairly close to our numerical results.
Hence it seems that that most suitable regular lattice for studying first
order phase transitions in two dimensions is a triangular lattice and
in three dimensions is a face centered cubic lattice.

The rather low value of $p_c$ in three dimensions means that domain
walls formed between degenerate vacua ($p=0.5$) will percolate and
almost all of the wall energy will be in one infinite
wall. Furthermore, even if the vacua are not degenerate, {\it i.e.}
there is bias in the system, infinite domain walls can still be produced.
If the properties of the bubble lattice are insensitive to small biases,
infinite domain walls would be produced for $p \gtrsim 0.17$. However,
it is likely that the bubble lattice will depend on the bias in three
ways. First, the nucleation rate of bubbles of the metastable vacuum
will be suppressed compared to that of the true vacuum. Secondly, the
velocity with which bubbles of the two phases grow can be different. Thirdly,
bubbles of the true vacuum may nucleate within the metastable vacuum.
In addition to these factors, bubbles may not retain their spherical
shape while expanding due to instabilities in their growth.
The effect of these factors on the percolation probability will be
model dependent. For example, the bubble velocities will depend on the
ambient plasma, and the nucleation rates on the action of the instantons
between the different vacuua. The inclusion of all these effects is
beyond the scope of the paper. Even so, a precise estimate of the
percolation probability in certain particle physics models will be 
extremely useful for studying cosmological scenarios such as described 
in Ref. \cite{gdhltv}. 

We have also investigated the formation of topological strings on the
random bubble lattice following the algorithm described in 
Ref. \cite{tvav}. We find that about 85\% of the strings in the 
simulation are infinite. This number should be compared with earlier
static simulations of string formation which yield a slightly lower
fraction ($\lesssim 80\%$). All these algorithms neglect phase equilibration 
processes when
domains of different phases collide and may be justified if the 
time scale $\tau$ is short compared to the typical time required for 
phase equilibration. In the case of domain walls, phase equilibration
in two colliding bubbles can only occur by the motion of the
phase separating wall across the volume of one of the bubbles. In
this case, the neglect of phase equilibration is justified if the domain
wall velocity is much smaller than the bubble wall velocity.

Finally, our analysis of the bubble lattice has an 
application to the large-scale structure of the universe. Astronomical
surveys show a universe filled with vast empty regions (voids) that
are outlined by walls of galaxies. If we place vertices at the centers
of the voids and then connect the vertices belonging to neighboring
voids, we will get a lattice much like those we have been considering
in the context of first order phase transitions.  Since the relation
(\ref{yzrelation}) is purely topological, it will also hold for the
dual void lattice. Further, working in the mean field approximation 
for the sizes of the voids, our estimate for ${\bar y}$ in (\ref{yestimate})
holds. Therefore we predict that a void should have 13.4
neighboring voids on average. This prediction will be modified if
there is spatial curvature in the universe, or if there are
correlations between the locations of void centers and their
sizes. (For example, if large voids are preferentially surrounded by
small voids.)  The curvature modification is, however, proportional to
$\chi^2$ where $\chi \lesssim 10^{-2}$ is the void size divided by the
spatial curvature radius, and hence is negligible for cosmology.

The modification in the void coordination number due to other 
cosmological factors such as cosmic expansion needs to be investigated 
further but a significant effect may be turned around to provide a probe 
of large-scale structure formation. Indeed, the distribution of coordination 
number (corresponding to Fig. \ref{neighbor}) may turn out to be a valuable 
tool in characterizing the large-scale structure. For example, structure 
formation scenarios based on topological defects are likely to yield different
results. In the specific case of the cosmic string 
scenario \cite{tv86,haramiyoshi}, voids 
form on either side of a string wake and filaments form where two
string wakes intersect, and so four (and not three) voids will neighbor
a filament. This will lead to a void lattice that is not
triangulated which will result in a lower mean coordination number.
Note that the difficulty associated with defining the boundary of a void 
does not enter the distribution of coordination number because the number 
of neighbors of a given void is insensitive to the precise location of 
the boundaries of the voids.

The large-scale structure has often been compared to a Voronoi foam 
\cite{icke}. However, the phase transition model appears to be more
suitable since two cells of the Voronoi foam stop growing in 
the direction of their collision, whereas this is not the 
behavior expected of large-scale voids. 
When two voids collide, they are better modelled
as if they continue to grow as in the case of bubbles in a phase
transition. Galaxies may be assumed to form in the space between
bubbles which will be sheetlike while the bubbles have not collided.
Upon collision, the sheetlike distribution of galaxies will get punctured.
The mutual collision of three voids will yield filamentary structure
and that of four voids will produce point-like structure.
From these considerations, the growth of voids is more like the growth 
of bubbles than of crystals for which the Voronoi and Johnson-Mehl 
models are applicable, and the universe is more like a system currently 
undergoing a first-order phase transition than like one that is crystallizing. 
One can also consider refinements of the phase transition model that would 
make it yet more like the evolution of large-scale structure. For example, 
the growth rate of voids will be time dependent, though they would nucleate
at the same epoch. Also, one could include processes in which a small 
void gets subsumed by a large void. We do not consider these details
further in this paper.

In conclusion, we have shown that the bubble distribution resulting from a
first order phase transition has a universal character and has a coordination
number that we have determined analytically. The analysis shows that a
triangular lattice in two dimensions and a face centered cubic lattice
in three dimensions are the regular lattices that come closest to
the bubble lattice. The study of percolation on the bubble lattice also 
supports this finding and we have found the critical percolation probability 
for the bubble lattice in both two and three dimensions. The result shows that 
infinite domain walls and strings will be produced in three dimensions. 
Finally we have applied our results to the void lattice in the 
large-scale structure of the universe and, based on some general 
assumptions, predict an average of 13.4 neighbors to a void. 

\

\noindent{\bf Acknowledgments}

We are grateful to Nick Rivier for his crucial input and to Harsh Mathur 
for extensive discussions throughout the course of this work. Comments
by Christopher Thompson are gratefully acknowledged. 
This research was supported by the DoE.


\end{document}